# Quantum anomalous Hall multilayers grown by molecular beam epitaxy

Gaoyuan Jiang[1,#], Yang Feng[1,#], Weixiong Wu[1], Shaorui Li[1], Yunhe Bai[1], Yaoxin Li[1], Qinghua Zhang[2], Lin Gu[2], Xiao Feng[1], Ding Zhang[1], Canli Song[1], Lili Wang[1], Wei Li[1], Xu-Cun Ma[1](, Qi-Kun Xue[1], Yayu Wang[1]*, Ke He[1]*

[1] State Key Laboratory of Low-Dimensional Quantum Physics, Tsinghua University, Beijing 100084
[2] Beijing National Laboratory for Condensed Matter Physics, Institute of Physics, Chinese Academy of Sciences, Beijing 100190

[#] These authors contributed equally to this work.
∗ Email: yayuwang@tsinghua.edu.cn; kehe@tsinghua.edu.cn

**Abstract** Quantum anomalous Hall (QAH) effect is a quantum Hall effect that occurs without the need of external magnetic field. A system composed of multiple parallel QAH layers is an effective high Chern number QAH insulator and the key to the applications of the dissipationless chiral edge channels in low energy consumption electronics. Such a QAH multilayer can also be engineered into other exotic topological phases such as a magnetic Weyl semimetal with only one pair of Weyl points. This work reports the first experimental realization of QAH multilayers in the superlattices composed of magnetically doped $(Bi,Sb)_2Te_3$ topological insulator and CdSe normal insulator layers grown by molecular beam epitaxy. The obtained multilayer samples show quantized Hall resistance $h/Ne^2$, where $h$ is the Planck's constant, $e$ is the elementary charge and $N$ is the number of the magnetic topological insulator layers, resembling a high Chern number QAH insulator.

A quantum anomalous Hall (QAH) insulator (or Chern insulator) possesses topologically non-trivial band structure characterized by a non-zero Chern number (*C*). At each edge of a QAH insulator, there are *C* dissipationless chiral edge channels which give rise to quantized Hall resistance $h/Ce^2$ (*h* is the Planck constant and *e* is the elementary charge) without the need of external magnetic field.[1,2] The dissipationless edge channels of QAH insulators have long been proposed to be used as interconnects in integrated circuits to reduce the enormous energy consumption in massive data transfer.[3] This concept has become realistic since the experimental realization of QAH effect in a magnetically doped $(Bi,Sb)_2Te_3$ topological insulator (TI).[4] So far, all the experimentally achieved QAH materials have a Chern number *C*=1, namely, with only one chiral edge channel at each edge.[4-9] The applications in low energy-consuming interconnects prefer QAH systems carrying as many edge channels as possible. The reason is that although electron transport on a chiral edge channel is dissipationless, dissipation does occur at its contacts to the current electrodes which are known as "hot spots".[10,11] The contact resistance at the hot spots cannot be reduced to below $h/e^2$, a limitation set by the quantum transport theory.[12] The resistance limit is inversely proportional to the number of parallel chiral edge channels between the electrodes and therefore can be reduced in high Chern number QAH insulators. However, a high Chern has not been realized in experiment so far. It is possible to obtain high Chern number

QAH phase in a magnetically doped (Bi,Sb)$_2$Te$_3$, but it would require a complex band structure and a smaller magnetic gap which means a lower QAH temperature.[11]

A more straightforward way to construct a high Chern number QAH system is by stacking several $C = 1$ QAH films into a multilayer structure decoupled by a normal (topologically trivial) insulator film inserted between adjacent QAH layers. Actually, such a multilayer not only acts as an effective high Chern number QAH insulator but probably hosts other exotic topological phases. For example, in a QAH multilayer, by tuning the inter- and intra-layer electronic coupling, one can drive it into a magnetic Weyl semimetal carrying one pair of Weyl points.[13] Coulomb interaction might induce novel quantum phenomena in a QAH bilayer, just as it induces exciton condensation and Coulomb drag in quantum Hall bilayers.[14,15] The rich phases and quantum phenomena make QAH multilayers a unique platform to study topological states of matter and their applications.

Though simple in concept, a QAH multilayer showing high order QAH effect has never been achieved before. The biggest challenge lies in the choice of material for the normal insulator spacing layers. The material must be a topologically trivial insulator with a large bandgap and negligible conductivity. It must have good lattice-match with the QAH material so that it can grow into a single-crystalline film on a QAH layer and support the epitaxial growth of another QAH layer. The bandgaps of the normal insulator and the QAH insulator must be well aligned without a conductive space charge layer forming at the interface. In this study, we found that CdSe(001) films satisfy all these requirements and can form QAH multilayers with magnetically doped (Bi,Sb)$_2$Te$_3$ films.

CdSe is a II-IV semiconductor with bulk gap of ~1.74 eV. Its stablest phase has the wurtzite structure as shown in **Figure 1**a.[16] In the hexagonal (001) plane, the lattice constant is 4.30 Å, which is between those of Bi$_2$Te$_3$ (4.43 Å) and Sb$_2$Te$_3$ (4.26 Å) and very close to that of magnetically doped (Bi,Sb)$_2$Te$_3$ QAH films. We grow CdSe(001) films with molecular beam epitaxy (MBE) on magnetically doped (Bi,Sb)$_2$Te$_3$ QAH films by directly evaporating CdSe compound (with purity 99.999%) with a commercial Knudsen cell. During the growth, the substrate is kept at 200 ℃—a temperature high enough to form a single-crystalline CdSe film but low enough to avoid the degrading of the QAH film.[9] The reflection high-energy electron diffraction (RHEED) patterns of a 6 quintuple (QL) (Cr$_{0.02}$V$_{0.16}$)(Bi$_{0.34}$Sb$_{0.66}$)$_{1.82}$Te$_3$ (CVBST) film and a CdSe film grown on it are shown in Figures 1b and 1c, respectively. The CdSe film shows streaky diffraction spots at exactly the same positions as those of the CVBST film, which indicates the formation of a two-dimensional (2D) CdSe film epitaxied on the QAH film.

Figures 1d and 1e display the angle-resolved photoemission spectroscopy (ARPES) bandmap and energy distribution curves of the CdSe film, respectively. Around the Fermi level ($E_F$) there is a clear bandgap, and the valence band maximum is ~1.3 eV below $E_F$. Considering that the bandgap of CdSe is ~1.74 eV, the conduction band minimum should be ~0.44 eV above $E_F$. Thus the MBE-grown CdSe film is insulating. No detectable shift in the CVBST bands is observed with the CdSe deposition, which suggests small band bending near the interface. Based on the ARPES data, the band diagram of CVBST/CdSe heterostructure is plotted in Figure 1f. Clearly, CdSe layer does not introduce conduction channels to the system. To test how insulating the MBE-grown CdSe films are, we measured the resistances of a 15 nm thick CdSe film epitaxied on a 1 QL $Sb_2Te_3$ film on $SrTiO_3$ (111) (the 1 QL $Sb_2Te_3$ film is an insulating buffer layer used to support the epitaxial growth of CdSe) and a 15 nm thick CdSe film epitaxied on InP (111) substrate. In both samples, the resistance is above 200 MΩ at room temperature, which is insulating enough to be used as the spacing layers in QAH multilayers.

By repeatedly growing 6 QL (~6 nm) CVBST film and 3.5 nm CdSe film with MBE, we are able to fabricate superlattice structures between the two materials. **Figure 2**a and 2b show the schematic structure and a large-scale cross-sectional transmission electron microscope (TEM) image of a superlattice sample including four CVBST (6 nm) layers and three CdSe (3.5 nm) layers, respectively. The superlattice structure is clearly seen in the TEM image. Figure 2c exhibits a zoom-in TEM image near a CVBST/CdSe interface. One can identify the characteristic quintuple-layer lattice structure of CVBST. The wurtzite structure of CdSe is also observed except for the first unit cell adjacent to CVBST which shows a zincblende-like lattice. The stacking fault always occurs in the first one or two unit cells of CdSe films grown on CVBST, which might result from the strain effect. The interface between CVBST and CdSe is atomically sharp and has nearly perfect epitaxy.

We also used x-ray diffraction (XRD) to characterize the crystalline structure of a similar multilayer sample with CdSe thickness about 6 nm. From the $2\theta$ diffraction curve, both the diffraction peaks of CVBST and CdSe are identified. At some of the diffraction peaks of CdSe or CVBST, satellite peaks are observed. From the zoom-in view of the CdSe (002) peak (inset of Figure 2d), one can see that it includes 4 satellite peaks, which are the signature of a superlattice structure with good periodicity over large scale. From the positions of the satellite peaks we can obtain the thickness of a CVBST/CdSe period $d \approx 12$ nm, consistent with the value estimated from the growth parameters.

**Figures 3**a-3d show the magnetic field dependences of the Hall ($R_{yx}$) and longitudinal ($R_{xx}$) resistances of CVBST (6 nm)/CdSe (3.5 nm) superlattice samples grown on $SrTiO_3$ with the number of CVBST layers ranging from 1 to 4. All measurements were carried out at 30 mK in a dilution refrigerator. A gate voltage is applied at the bottom of the $SrTiO_3$ substrate ($V_g$) to tune the chemical potential of each film such that $R_{yx}$ reaches the maximum and $R_{xx}$ reaches the minimum (the $N = 4$ sample was measured without gate voltage applied due to failure of the gate electrode.). The $V_g$ dependences of $R_{yx}$(0T) (blue) and $R_{xx}$(0T) (red) of the $N = 1$, 2 and 3 samples at zero magnetic field are displayed in Figures 3e-3g, respectively. For all the samples, the $R_{yx}$-$\mu_0H$ curves (blue ones in Figures 3a-3d) show rectangular hysteresis loops. The $R_{xx}$-$\mu_0H$ curves (red ones in Figures 3a-3d) exhibit two peaks at the coercive fields; elsewhere $R_{xx}$ is basically independent of $H$. All the samples have a similar coercivity of ~0.8 T. The abruptness of magnetization switching, as shown by the widths of the $R_{xx}$ peaks at coercive fields, is also similar in different samples. These observations indicate that CdSe layers have little influence on the magnetic properties of CVBST films. The anomalous Hall resistance ($R_{AH} = R_{yx}(0T)$) exhibits a systematic decrease with increasing number of the CVBST layers ($N$) (Figure 3(h)). In the $N = 1$ sample, $R_{AH}$ at zero magnetic field is $h/e^2$; in the $N = 2$ sample, $R_{AH}$ is nearly $h/2e^2$; in the $N = 3$ and $N = 4$ samples, $R_{AH}$ becomes $h/3e^2$ and $h/4e^2$, respectively. Clearly, the decrease of $R_{AH}$ with increasing number of QAH layers basically follows the relation $R_{AH}(N) = h/Ne^2$. As shown in Figures 3e-3g, the $R_{yx}(0T)$-$V_g$ curve at zero field always shows a plateau at or close to the quantized value $h/Ne^2$ which is always accompanied by a suppression in $R_{xx}$. The inverse correlation of $R_{yx}$ and $R_{xx}$, as well as the much larger value of the former than the latter, strongly indicates that the observed quantized $R_{yx}$ plateaus ($h/Ne^2$) are contributed by the QAH edge states. We thus conclude that the samples have $N$ chiral edge channels in parallel and are equivalent to a QAH insulators of Chern number $C = N$.

The QAH multilayer samples obtained so far are not perfect, as shown by the non-zero $R_{xx}$ corresponding to the quantized plateau. Similar to usual quantum Hall systems, $R_{xx}$ of a QAH system is more sensitive to sample quality than $R_{yx}$ and can easily deviate from zero. We can see from Figure 3h that $R_{xx}$ varies randomly in different samples. Although MBE-grown CdSe films are by and large insulating as demonstrated by the ARPES and transport measurements, conduction channels induced by chemical potential fluctuations and defects may exist and induce dissipation in the system. The problem can be solved by further improvement of the MBE growth conditions. Even in the present samples, the total resistance $R_{tot} = \sqrt{R_{xx}^2(0T) + R_{yx}^2(0T)}$, which depicts the whole sample dissipation including that at the hot spots, obviously decreases with the increasing number of QAH layers (Figure 3h). It demonstrates that QAH multilayers can indeed solve the hot spot problem for the applications of chiral edge states as interconnections. One can in principle make the dissipation as low as possible by simply increasing the number of QAH layers.

The QAH multilayers provide a way to realize a magnetic Weyl semimetal with only one pair of Weyl points. A Weyl semimetal has energy bands including one or several pairs of Weyl points where there is neither bandgap nor density of states, and the spin texture nearby exhibits magnetic monopoles in momentum space. There are two kinds of Weyl semimetals: one with broken spatial inversion symmetry, and the other with broken time-reversal symmetry.[17] The former one has been found in several materials, all of which have many pairs of Weyl points and thus physical properties are difficult to understand.[18-22] So far, there is no convincing experiment demonstrating the latter case yet, in spite of many predictions.[23-25] The time-reversal-symmetry-broken, or magnetic, Weyl semimetal can be fabricated with a bottom-up approach: by growing a superlattice structure of magnetic topological insulator and normal insulator layers.[13] According to the theoretic phase diagram of such a superlattice structure, one can drive a QAH multilayer into magnetic Weyl semimetal phase by enhancing the hybridizations of the topological surface states of different layers when reducing the thicknesses of both the QAH and normal insulator layers. The realization of CVBST/CdSe QAH multilayers thus paves the road for the superlattice approach to a magnetic Weyl semimetal. Furthermore, this Weyl semimetal has only one pair of Weyl points, much simpler than the existing Weyl semimetal materials, which can be used as a model Weyl semimetal system to demonstrate its unusual properties.

To conclude, using CdSe films as spacing layers, we successfully fabricated multilayer structures of magnetically doped $(Bi,Sb)_2Te_3$ QAH insulators with MBE technique. The multilayers show the high order QAH effect with the total resistance significantly reduced. The QAH multilayers not only demonstrate a way to reduce the contact dissipation of QAH edge states in applications, but also provide an excellent platform to investigate some important issues on high order QAH effect, Weyl semimetal, correlation effects of chiral edge states, etc.

**Acknowledgements**

This work is financially supported by the National Key Research and Development Program of China (grant no. 2017YFA0303303), the National Natural Science Foundation of China (grant no. 51661135024), and the Beijing Advanced Innovation Center for Future Chip (ICFC).

**Figures and Figure Captions**

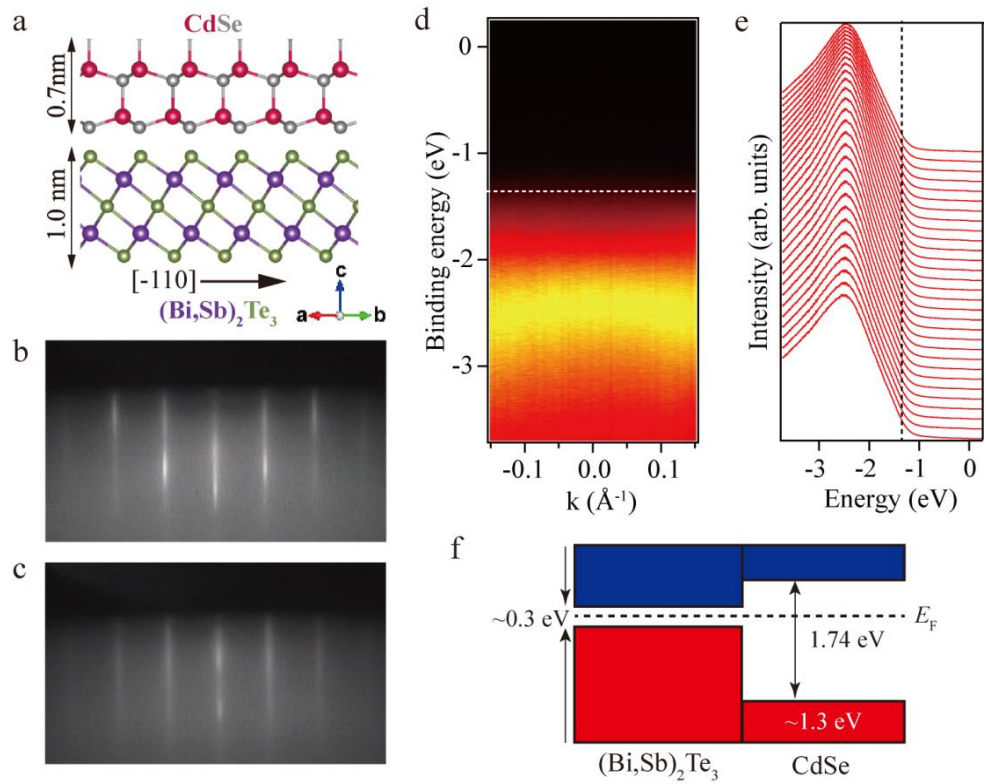

**Fig. 1.** MBE-grown CdSe on magnetic TI. a) Schematic lattice structure of CdSe and $(Bi,Sb)_2Te_3$. The direction of the cross section is [-110]. b), c) RHEED patterns of CVBST (b) and CdSe grown on CVBST (c). d), e) ARPES bandmap (d) and energy distribution curves (e) of 3.5 nm CdSe grown on 6 QL CVBST. The data are taken around $\bar{\Gamma}$ point along the $\bar{\Gamma}$ - $\bar{K}$ direction.

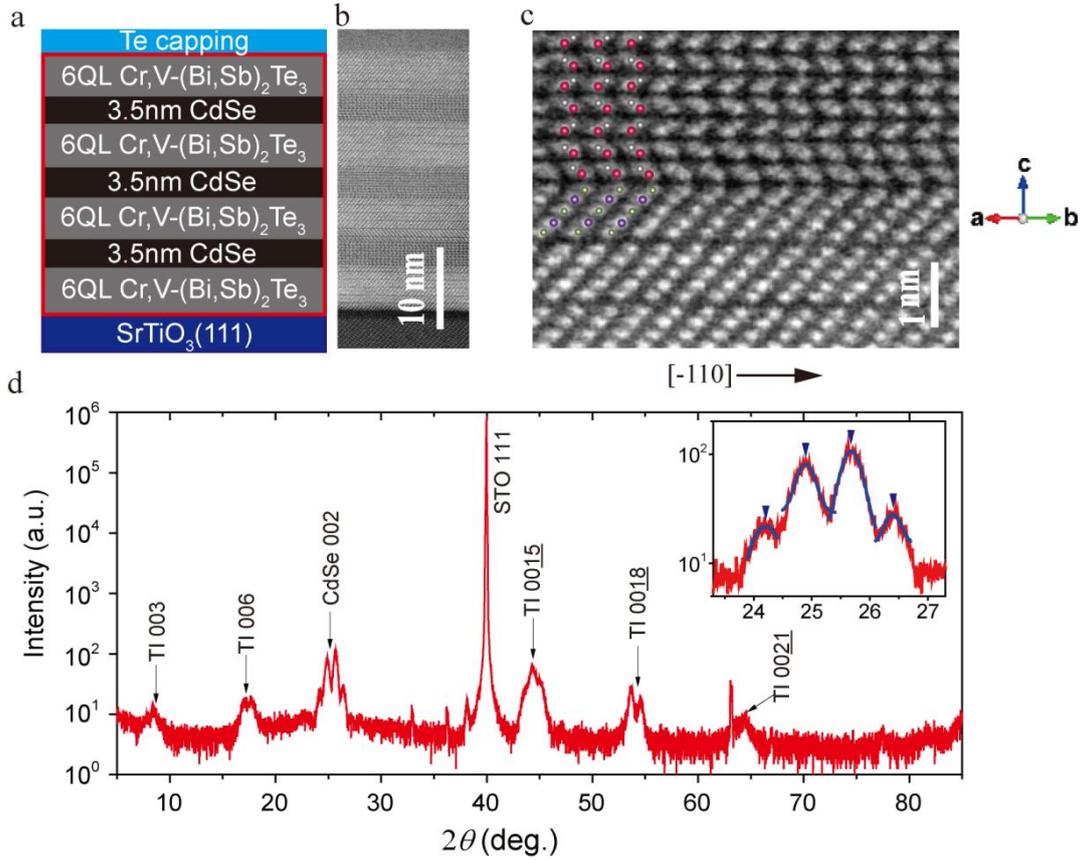

**Fig. 2.** Structural characterizations of CVBST/CdSe superlattice. a) Schematic structure of a CVBST/CdSe superlattice sample. b) Large-scale cross-sectional TEM image of the CVBST/CdSe superlattice. c) Zoom-in TEM image near the interface between CVBST and CdSe layers. The circles indicate the positions of atoms. d) X-ray diffraction pattern of a CVBST/CdSe superlattice. The inset shows a zoom-in view of the CdSe (002) peak with satellite peaks. Dark blue lines are Gaussian fitting of each satellite peak.

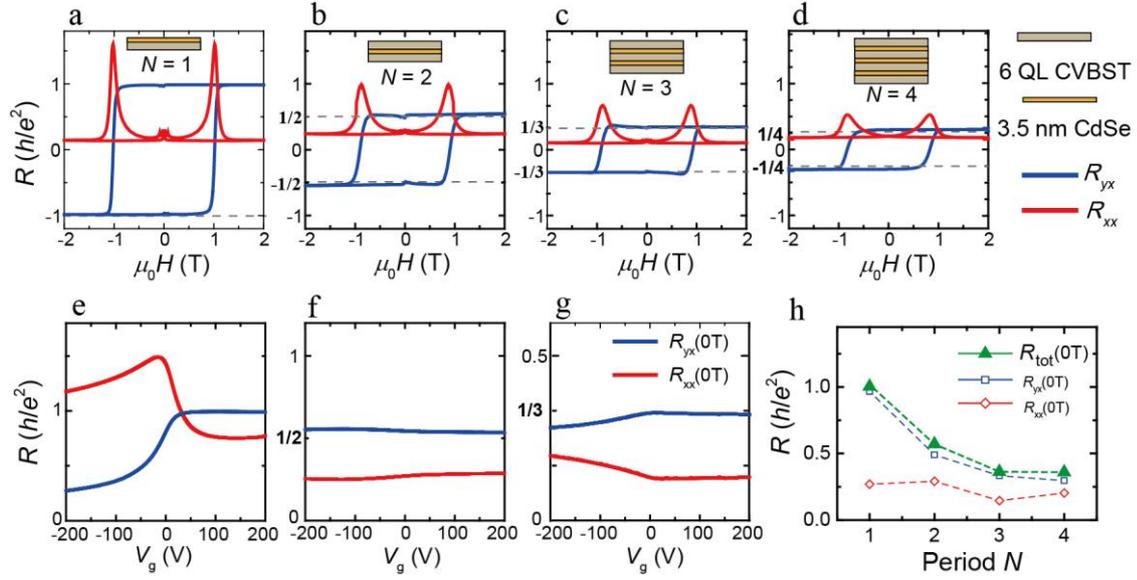

**Fig. 3.** Quantum anomalous Hall effect observed in [CVBST/CdSe]$_N$ superlattice samples at 30 mK. a)-d) Magnetic field ($\mu_0 H$) dependences of the Hall ($R_{yx}$, blue) and longitudinal ($R_{xx}$, red) resistances of the $N$ = 1 (a), 2 (b), 3 (c), and 4 (d) samples. e)-g) Bottom-Gate-voltage ($V_g$) dependences of the Hall ($R_{yx}$(0T), blue) and longitudinal ($R_{xx}$(0T), red) resistances of the $N$ = 1 (e), 2 (f), and 3 (g) samples at zero magnetic field. h) Dependences of Hall ($R_{yx}$(0T), blue), longitudinal ($R_{xx}$(0T), red), and total ($R_{tot} = \sqrt{R_{xx}^2(0\,\mathrm{T}) + R_{yx}^2(0\,\mathrm{T})}$) resistances on the number of CVBST layers.